\documentclass[]{rsos}

\usepackage{graphicx}   
\usepackage{xcolor} 
\begin{document} 
\title{An empirical study of the per capita yield of 
science Nobel prizes: Is the US era coming to an end?}

\author{Claudius Gros$^{1}$}
\address{$^{1}$Institute for Theoretical Physics, Goethe University Frankfurt, 
Frankfurt a.\,M., Germany}

\subject{e-science, complexity}

\keywords{Nobel prizes, predictive modeling, science of sciences}

\corres{Claudius Gros\\
\email{gros07[at]itp.uni-frankfurt.de}}

\begin{abstract} 
We point out that the Nobel prize production of the USA, the UK, Germany and
France has been in numbers that are large enough to allow for a reliable
analysis or the long-term historical developments. Nobel prizes are often
split, such that up to three awardees receive a corresponding fractional prize.
The historical trends for the fractional number of Nobelists per population are
surprisingly robust, indicating in particular that the maximum Nobel
productivity peaked in the 1970s for the US and around 1900 for both France and
Germany. The yearly success rates of these three countries are to date of the
order of 0.2-0.3 physics, chemistry and medicine laureates per 100 million
inhabitants, with the US value being a factor 2.4 down from the maximum
attained in the 1970s. The UK managed in contrast to retain during most of the
last century a rate of 0.9-1.0 science Nobel prizes per year and per 100
million inhabitants. For the USA one finds that the entire history of science
Noble prizes is described on a per capita basis to an astonishing accuracy by a
single large productivity boost decaying at a continuously accelerating rate
since its peak in 1972.  \end{abstract}

\begin{fmtext}    

\vspace{-1ex}
\section{Introduction}

The `science of sciences' has emerged over the last 
decades as a vibrant research field \cite{zeng2017science}, in
particular because it may open a route to `predict' discoveries
in research fields that have achieved a certain degree of
maturity \cite{clauset2017data}. On an individual level,
the investigations are focused mainly on measures quantifying 
the scientific excellence of a given 
scientist \cite{hirsch2005index,wang2013quantifying},
as well as the future impact of a research publication
\cite{ke2015defining,mukherjee2017nearly}. It remains
however a challenge to predict at which stage of her 
or his career a scientist will be likely to publish a 
breakthrough paper, if ever
\cite{sinatra2016quantifying,jones2011age,way2017misleading}.

\end{fmtext}
\maketitle

The situation changes when it comes to the overall performance 
of a country and its scientific institutions \cite{zeng2017science},
which is determined not by the achievements of individuals,
but by aggregate and hence averaged variables. The same
holds for the history of Nobel prizes in the natural sciences
for larger countries \cite{nobelCommittee}. Nobel prizes have 
been acquired by the USA, the UK, Germany and France at rates
that are steady enough, as we will show here, that the time 
evolution of the respective Nobel prize productivity can 
be both modeled accurately and forecast for the years to 
come. Such an analysis allows for an improved understanding 
of the history of scientific discoveries. It also provides
science managers insights on how the aggregate productivity 
of the scientific institutions of a country is developing.

One can model empirical data either via a straightforward fit
or by approximating it by an underlying model \cite{luduena2013large}.
An example of the first approach is the linear increase
in record human life expectancy that has been observed to 
hold for more than a century \cite{oeppen1029}. It is however 
difficult to gauge how long a trend found empirical will 
continue to persist. The historical raise of human life expectancy, 
to come back to this example, has been predicted repeatedly
to level off \cite{oeppen1029}, yet it keeps going \cite{gros2012pushing}.
The underlying drivings in terms of continuing progress in the medical 
sciences seem to remain operative.

Here we propose that the history of science Nobel
prize success can by analyzed by a sociophysical model 
describing two fundamental drivings, a underlying
long-term productivity rate and an extended but
otherwise temporary burst in science productivity.
We believe that the trends resulting from this model
for the next decade are relatively robust. Our results,
in particular that the US per capita science productivity
in terms of Nobel prizes in the natural sciences is
continuing to decline, warrant a critical discussion.

\begin{figure}[t]
\centerline{
\includegraphics[width=0.70\columnwidth]{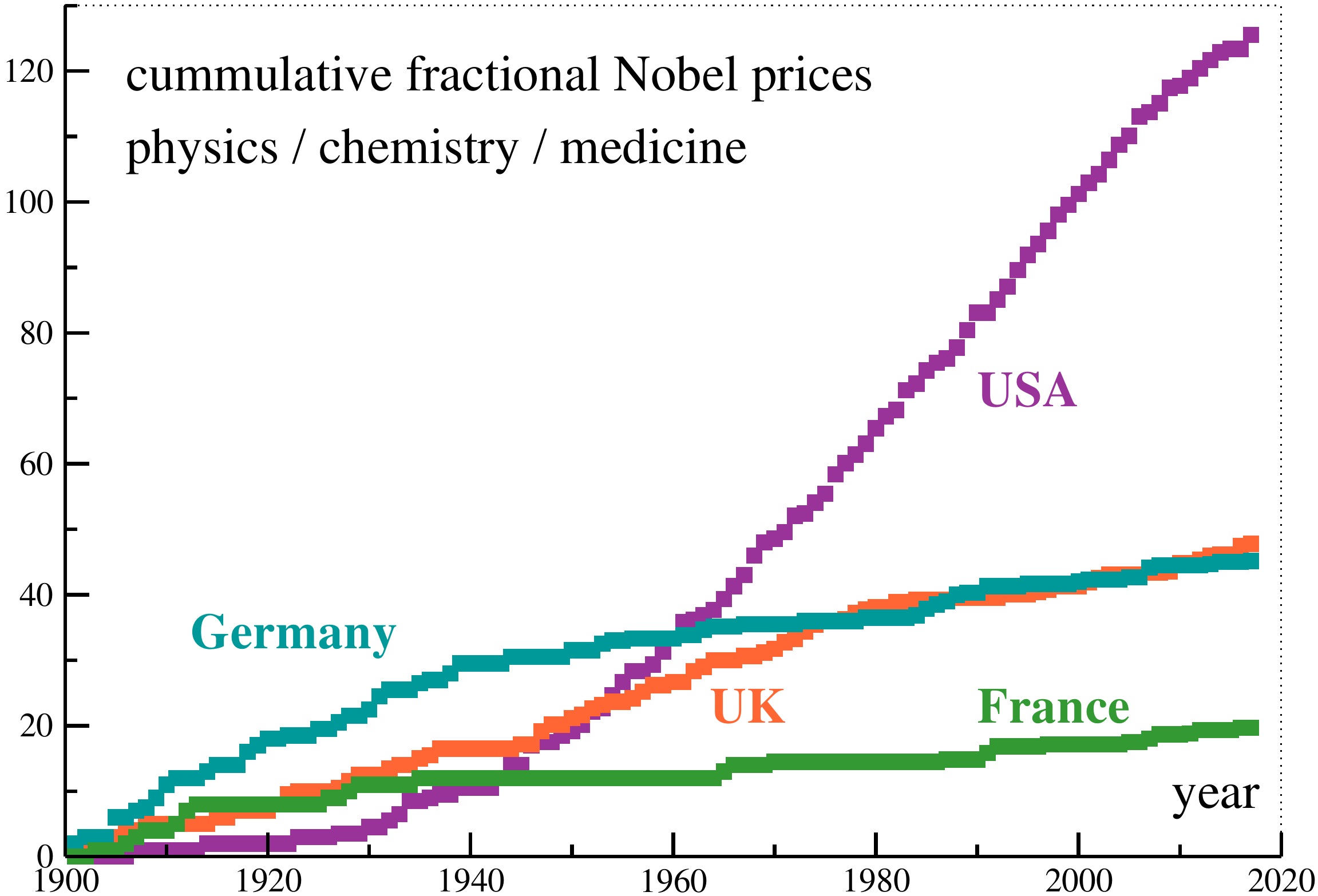}
           }
\caption{The cumulative per country number of physics, 
chemistry and medicine Nobel prizes. Prizes are 
attributed to the respective country according to the 
nationality the recipients had at the time of the 
announcement, with prizes obtained by more than one 
recipient accordingly divided. Note that the US population
increased from 76 to 327 million from 1901 to 2017.
}
\label{fig_bare}
\end{figure}

\section{Results} 

Nobel prizes are awarded as fractional prizes to up to 
three scientists. In Fig.\,\ref{fig_bare} we have plotted
the cumulative number of prizes, where the number of 
fractional laureates received in a given year was determined
according to the nationality of the recipients at the time of 
the prize announcement (modulo multiple citizenships, 
when present). The USA has been leading unsurprisingly ever 
since the 1960s. In order to determine the relative
contributions of the population size and of the excellence 
of the scientific institutions with regard to Nobel 
productivity, one needs to factor out the population 
size. This is particularly important when the population 
has been varying substantially. For USA, to give an 
example, the population increased from 76 million in 1901, 
the year the Nobel prize was first awarded, to the present 
day value of 327 million.

\begin{figure}[t]
\centerline{
\includegraphics[width=0.70\columnwidth]{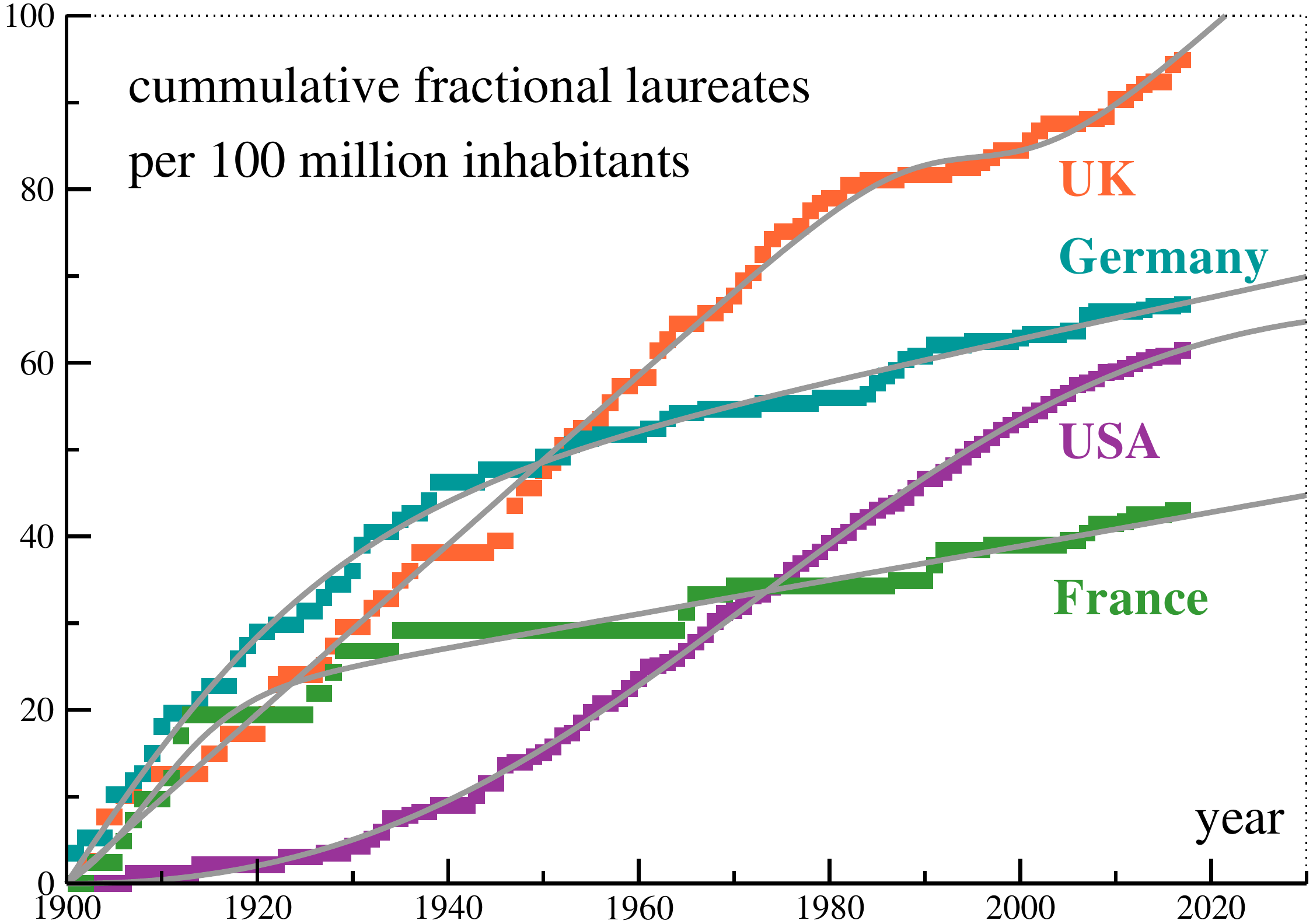}
           }
\caption{Science Nobel prizes per 100 million inhabitants.
The historical population data at the time of the announcement
where used and the such obtained yearly increments cumulatively 
added. The data can be modeled (grey lines, see Eq.~(\ref{eq_model}))
by a superposition of a linear growth term and a one time 
period of either increased (as for the USA, Germany and France)
or reduced productivity (as for the UK), centered respectively
around 1898, 1909, 1972 and 1995 for Germany, France, the USA
and UK.
}
\label{fig_perPop}
\end{figure}

For the yearly per capita yield we divided the fractional 
number of Nobel prizes a country received in a given
year by the population the country of nationality had 
in the same year, interpolating whenever necessary the 
respective census data 
\cite{UScensus,UKcensus,GERMANYcensus,FRANCEcensus}.
The yearly counts obtained in this manner were added subsequently 
to a cumulative measure that has the advantage of fluctuating
substantially less than the yearly increments. The resulting 
data is presented in Fig.\,\ref{fig_perPop}, where we used 
a reference population of 100 million inhabitants. On a per 
capita basis the winner is the UK, with Germany coming second
and the US a close third. Prolonged periods without Nobel
prizes show up in plateaus.

\subsection*{Modeling} 

The historical evolution of the per capita yield
of fractional Nobel laureates presented in Fig.\,\ref{fig_perPop} 
can be fitted by a model,
\begin{equation}
\gamma t+\alpha g\left(\frac{t-t_{max}}{\tau_{max}}\right)-c_0\,,
\label{eq_model}
\end{equation}
that incorporates a background of constant success
$\propto\gamma$. The second term in (\ref{eq_model}) 
is proportional to the S-shaped logistic function 
$g$, with its derivative, 
\begin{equation}
\frac{d g}{dx}\,=\, g(1-g),
\qquad\quad
g(x)\,=\,\frac{1}{1+\exp(-x)}\,
\label{eq_g_prime}
\end{equation}
corresponding either to a burst in productivity
centered around $t_{max}$, if $\alpha>0$, or to a period 
with decreased Nobel success when $\alpha<0$. The respective
decay time is $\tau_{max}$. The constant $c_0$ entering 
(\ref{eq_model}) ensures that the count starts at zero 
for $t\!=\!1900$, that is in the year before Nobel prizes 
where first awarded. 

The parameters entering the sociophysical model
(\ref{eq_model}) can be determined via a straightforward
least square fit. For $(t_{max},\tau_{max})$ one 
obtains (1909,5) for France, (1898,15) for Germany, 
(1995,5) for the UK and (1972,29) for the USA.
The corresponding results for $(\gamma,\alpha)$ 
are (-0.34,135) for the USA, (0.98,-18.8) for
the UK, (0.24,84) for Germany and (0.2,22.4)
for France. 

The set of parameters contains two particular cases. 
The negative $\gamma<0$ implies for the USA that a
long-term productivity rate cannot be extracted,
viz that the boost $\sim\alpha$ completely dominates
the US history of science Nobel prizes. For the
UK one obtains that $\alpha<0$, which indicates
that the country experienced in the 1990s a phase of 
reduced (and not of increased) Nobel prize productivity.

\begin{figure}[t]
\centerline{
\includegraphics[width=0.70\columnwidth]{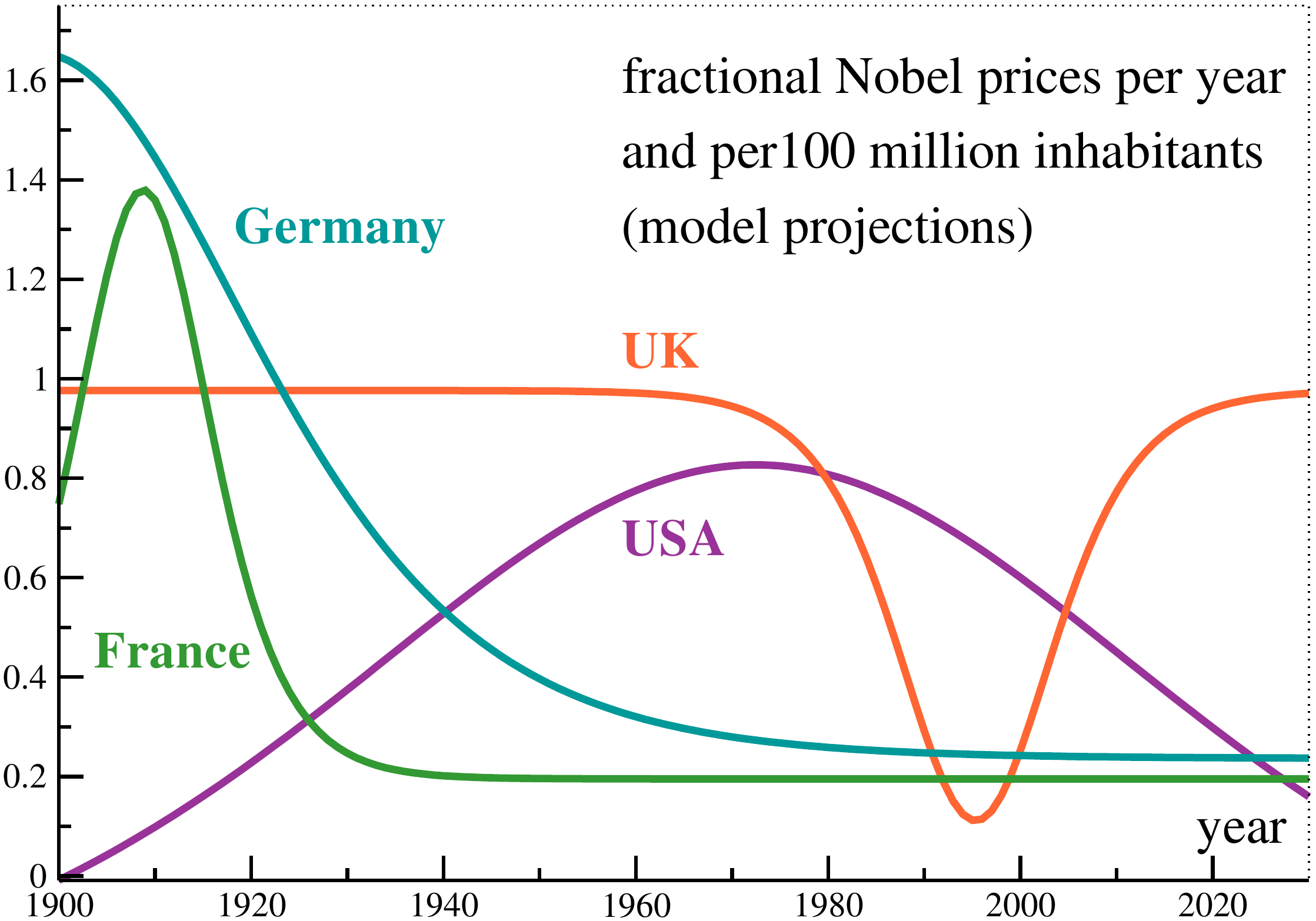}
           }
\caption{Historical Nobel productivity. The average number
of fractional science Nobel prizes received per year and per 
100 million inhabitants, as given by the derivative
$\gamma + \alpha g(1-g)/\tau_{max}$ of the respective analytic 
models. Compare Eq.~(\ref{eq_model}) and Fig.\,\ref{fig_perPop}.
}
\label{fig_productivity}
\end{figure}

The curves corresponding to (\ref{eq_model}),
which have been included in Fig.\,\ref{fig_perPop}, 
approximate the data to an astonishing degree, in 
particular for the USA. This observation suggests
that the derivative of the analytic model with 
respect to time, namely $\gamma + \alpha g(1-g)/\tau_{max}$, 
constitutes a reliable estimate for the evolving 
aggregate per capita Nobel prize productivity of
a given country. The result is presented in 
Fig.\,\ref{fig_productivity}. One finds that the
four countries discussed here are surprisingly diverse 
with respect to how the efficiency of their scientific 
institutions evolved over the last century in terms 
of per capita Nobel productivity.

\section{Discussion}

Our results show that countries may vary drastically
with respect to the historical development of their
science productivity. 

\subsection*{France} 

After a short boost in Nobel productivity that peaked in 1909,
France produced Nobel laureates in physics, chemistry and
medicine at a relative constant rate of 0.2 per year and 
per 100 million inhabitants. The overall number of fractional 
medals received from 1901 to 2017 is 20. One needs to
add that French scientific institutions excel furthermore 
in areas that do not affect the Nobel prize yield, in
particular in mathematics. 13/12/6/1 Fields medals were
won respectively by the USA/France/UK/Germany. With a
2017 population of 65 million, France has by far 
the highest per capita success rate.

\subsection*{Germany} 

Germany's science productivity peaked in 1898, which
antedates the first Nobel prize by three years. The Nobel
prize came hence somewhat too late for Germany, which would 
have received a substantially larger amount of medals if 
the first prize had not been awarded in 1901, but 20 years 
earlier. Today's rate is 0.24 science Nobel prizes per year 
and 100 million inhabitants. 

German science funding is focusing these days more and more 
on cooperative funding programs that are per se a response 
to the diminishing returns observed in many mature research 
fields \cite{gros2012pushing}. It is to been seen if the
concurring progressive marginalization of individual research 
will will have a detrimental impact, in the long run, on
science productivity at the highest level in terms of Nobel 
medals.

\subsection*{UK} 

The UK received during most of the last century 
a whopping 0.98 science Nobel prizes per year and
per 100 million inhabitants. This remarkable streak has 
been interrupted temporarily in the mid 1990s, but we 
caution that the value for the depth of this depression 
found by the analytic model is to be taken only as an 
order of magnitude estimate. It will be interesting to 
see whether the 1990s depression in the number of UK 
science Nobel prizes was a one-time event or whether 
it bodes rougher times ahead.

\subsection*{USA} 

The entire US history of science Nobel prizes can be 
interpreted in terms of a single large productivity boost 
peaking in 1972. The extraction of a long-term basic 
productivity rate, if such a rate should be present,
is preempted by the extraordinary long decay time 
of 29 years. A steady state with regard to scientific 
research has yet to be reached. 

The average number of fractional science Nobel prizes the US receives 
presently per year and per 100 million inhabitants is 0.34, a 
respectable value, which is however down by a factor 2.4 relative
to the peak value of 0.83 reached in 1972. Striking is moreover 
the continuing downward trend. Our model predicts that the US 
per capita productivity rate will have fallen below the one of 
Germany by 2025 and below the one of France by 2028. It is hard 
to imagine a scenario, given the remarkable accuracy of the 
analytic model, for which the final bottoming out of the US 
per capita productivity of science Nobel laureates would not 
occur at very low levels.

The decline of the US per capita productivity of science
Nobel laureates has been masked hitherto by the concurring 
increase of the US population, which equalled a factor of 
1.6 between 1972 and 2017 (from 208 to 327 million). The 
continuously growing population size allowed the US to increase 
the overall count of fractional science Nobel prizes in the 
same period from 52 to an impressive 126. 

One may argue that per capita levels are destined to fall in 
a world in which the overall population grows at an unabated 
pace and in which more countries than ever fund scientific 
research.  While undoubtedly true, this argument falls short 
to explain the large differences robustly observed when comparing 
France, Germany and the UK to the USA. Other factors must
hence determine the observed decline of the US per capita 
productivity of Nobel laureates in the natural sciences.

National science funding policies can be clearly successful
regardless of the prospect of acquiring Nobel prizes, in particular 
because Nobel prizes do not cover new fields like computer
science, a typical US domain \cite{bonaccorsi2017explaining}.
Is the ongoing decline of the US per capita physics, chemistry
and medicine Nobel prize success then cause for alarm or nothing
else than an indication for a paradigm shift, that is for a 
refocusing of research priorities towards new and more
rewarding areas? 

\section{Conclusion}

We have shown 
that the timeline of physics, chemistry and medicine 
Nobel prizes provides a reliable database that allows
to extract both a tractable model and to forecast future 
aggregate developments. We have pointed out in particular that
the USA per capita success rate is declining ever since 1972 
and that this terminal trend has been masked hitherto by the
concurring growth of the population. The resulting prediction, 
namely that this downward trend will continue for the decade 
to come, allows to either corroborate or to invalidate the model 
proposed here. Future investigations may take the time lag into 
account, that passes between a discovery and its honoring by a 
Nobel medal \cite{fortunato2014growing}.


\vskip1pc
\ethics{Not applicable.}

\dataccess{``This article has no additional data''.}

\aucontribute{All work done by C.Gros.}

\competing{Not applicable.}

\funding{Not applicable.}

\ack{The author thanks R.~Valent\'\i\ for discussions.}

\disclaimer{Not applicable.}



\begin{thebibliography}{99}

\bibitem{zeng2017science}
Zeng A, Shen Z, Zhou J, Wu J, Fan Y, Wang Y, Stanley HE. 2017  The science of
  science: From the perspective of complex systems. {\em Physics Reports}.

\bibitem{clauset2017data}
Clauset A, Larremore DB, Sinatra R. 2017  Data-driven predictions in the
  science of science. {\em Science} \textbf{355}, 477--480.

\bibitem{hirsch2005index}
Hirsch JE. 2005  An index to quantify an individual's scientific research
  output. {\em Proceedings of the National academy of Sciences of the United
  States of America} \textbf{102}, 16569.

\bibitem{wang2013quantifying}
Wang D, Song C, Barab{\'a}si AL. 2013  Quantifying long-term scientific impact.
  {\em Science} \textbf{342}, 127--132.

\bibitem{ke2015defining}
Ke Q, Ferrara E, Radicchi F, Flammini A. 2015  Defining and identifying
  Sleeping Beauties in science. {\em Proceedings of the National Academy of
  Sciences} \textbf{112}, 7426--7431.

\bibitem{mukherjee2017nearly}
Mukherjee S, Romero DM, Jones B, Uzzi B. 2017  The nearly universal link
  between the age of past knowledge and tomorrow's breakthroughs in science and
  technology: The hotspot. {\em Science Advances} \textbf{3}, e1601315.

\bibitem{sinatra2016quantifying}
Sinatra R, Wang D, Deville P, Song C, Barab{\'a}si AL. 2016  Quantifying the
  evolution of individual scientific impact. {\em Science} \textbf{354},
  aaf5239.

\bibitem{jones2011age}
Jones BF, Weinberg BA. 2011  Age dynamics in scientific creativity. {\em
  Proceedings of the National Academy of Sciences} \textbf{108}, 18910--18914.

\bibitem{way2017misleading}
Way SF, Morgan AC, Clauset A, Larremore DB. 2017  The misleading narrative of
  the canonical faculty productivity trajectory. {\em Proceedings of the
  National Academy of Sciences} p. 201702121.

\bibitem{nobelCommittee}
 The Nobel Committee. \url{http://www.nobelprize.org/}.

\bibitem{luduena2013large}
Ludue{\~n}a GA, Meixner H, Kaczor G, Gros C. 2013  A large-scale study of the
  world wide web: network correlation functions with scale-invariant
  boundaries. {\em The European Physical Journal B} \textbf{86}, 348.

\bibitem{oeppen1029}
Oeppen J, Vaupel JW. 2002  Broken Limits to Life Expectancy. {\em Science}
  \textbf{296}, 1029--1031.

\bibitem{gros2012pushing}
Gros C. 2012  Pushing the complexity barrier: diminishing returns in the
  sciences. {\em Complex Systems} \textbf{21}, 183.

\bibitem{UScensus}
 United States Census Bureau. \url{http://www.census.gov/}.

\bibitem{UKcensus}
 Office for National Statistics of the United Kingdom.
  \url{http://www.ons.gov.uk/}.

\bibitem{GERMANYcensus}
 Bundeszentrale f\"ur politische Bildung. \url{http://www.bpb.de/}.

\bibitem{FRANCEcensus}
 Institut national de la statistique et des \'etudes \'economiques.
  \url{https://www.insee.fr/}.

\bibitem{bonaccorsi2017explaining}
Bonaccorsi A, Cicero T, Haddawy P, Hassan SU. 2017  Explaining the
  transatlantic gap in research excellence. {\em Scientometrics} \textbf{110},
  217--241.

\bibitem{fortunato2014growing}
Fortunato S, Chatterjee A, Mitrovic M, Pan R, Parolo P, Beccattini F. 2014
  Growing time lag threatens Nobels. {\em Nature} \textbf{508}, 186--186.

\end{thebibliography}

\end{document}